# A Network Object Method to Uncover Hidden Disorder-Related Brain Connectome

Shuo Chen*[1], Yishi Xing[1], Jian Kang[2], Dinesh Shukla[3], Peter Kochunov[3], and L. Elliot Hong*[3]

[1] Department of Epidemiology and Biostatistics, University of Maryland, College Park, MD20742, USA

[2] Department of Biostatistics, University of Michigan, Ann Arbor, MI, 48109, USA

[3] Maryland Psychiatric Research Center, Department of Psychiatry, University of Maryland School of Medicine, Baltimore, MD, 21228, USA

## Abstract

Neuropsychiatric disorders impact functional connectivity of the brain at the network level. The identification and statistical testing of disorder-related networks remains challenging. We propose novel methods to streamline the detection and testing of the hidden, disorder-related connectivity patterns as network-objects. We define an abnormal connectome subnetwork as a network-object that includes three classes: nodes of brain areas, edges representing brain connectomic features, and an organized graph topology formed by these nodes and edges. Comparing to the conventional statistical methods, the proposed approach simultaneously reduces false positive and negative discovery rates by letting edges borrow strengths precisely with the guidance of graph topological information, which effectively improves the reproducibility of findings across brain connectome studies. The network-object analyses may provide insights into how brain connectome is systematically impaired by brain illnesses.

## Introduction

Neuropsychiatric brain disorders often involve systematic impairment of brain functional or/and structural connectome at the circuitry level[1-4]. Identification of impacted networks is necessary for understanding the neural pathophysiological mechanism and may lead to discovery of network-level biomarkers[5-9]. Neuroimaging techniques provide non-invasive ways to measure and quantify connections between brain areas of human subjects[1-10]. However, disease-relevant networks typically cannot be fully described prior to experiments, and thus are challenging to detect with statistical rigor[10-12]. To address this challenge, we propose a **n**etwork-**o**bject **s**tatistics (NOS) approach to extract and test hidden, disease-related brain connectome subnetworks.

To illustrate the concept and what the new method aims to achieve, we start with a simple synthetic data set as shown in Figure 1. Assume that a study has collected resting state functional magnetic resonance imaging (rs-fMRI) from 50 patients with a neuropsychiatric illness and 50 matched healthy controls. Twenty brain regions of interest (ROIs) are included as nodes in Figure 1. An edge connecting between two nodes represents their connectivity. The strength of the connectivity can be represented by a number of methods, for example, the coherence statistic of two time series (say a correlation coefficient) from two brain areas is used for functional connectivity (FC). The functional connectivity metrics (e.g. Fisher's Z transformed Pearson correlation coefficients of time series) in this case has 190 weighted edges between 20 ROIs for each subject. One then faces a multiple testing problem to compare the 190 connectivity metrics (edges) between the two cohorts. Figure 1a displays the negative log transformed p-value heatmap (a p-value is obtained from two-sample t test on each edge). Figure 1b shows node and edge representation where red edges indicate the (true) differentially expressed connectivity in patients compared with controls; blue edges for the (true) non-differentially expressed edges. However, it would be difficult for existing statistical methods to identify the differentially expressed connectomic networks with organized yet latent topological graph structures. The mass univariate methods including the family-wise error rate (FWER), the false positive discovery rate (FDR), and the network based statistics (NBS) could identify a set of differentially expressed *individual* edges but miss the opportunity to identify the underlying disease-relevant and organized graph topological structure (pathway)[2,12]. Similarly, the dimension reduction methods (e.g. the independent component analysis) could detect differential components but miss the underlying network structure. On the other hand, the descriptive graph metrics (e.g. small-worldness and modularity) could identify the differential global graph topological properties across groups, but not be able to localize nor recognize the organized graph topological structures of the differentially expressed connectomic networks.

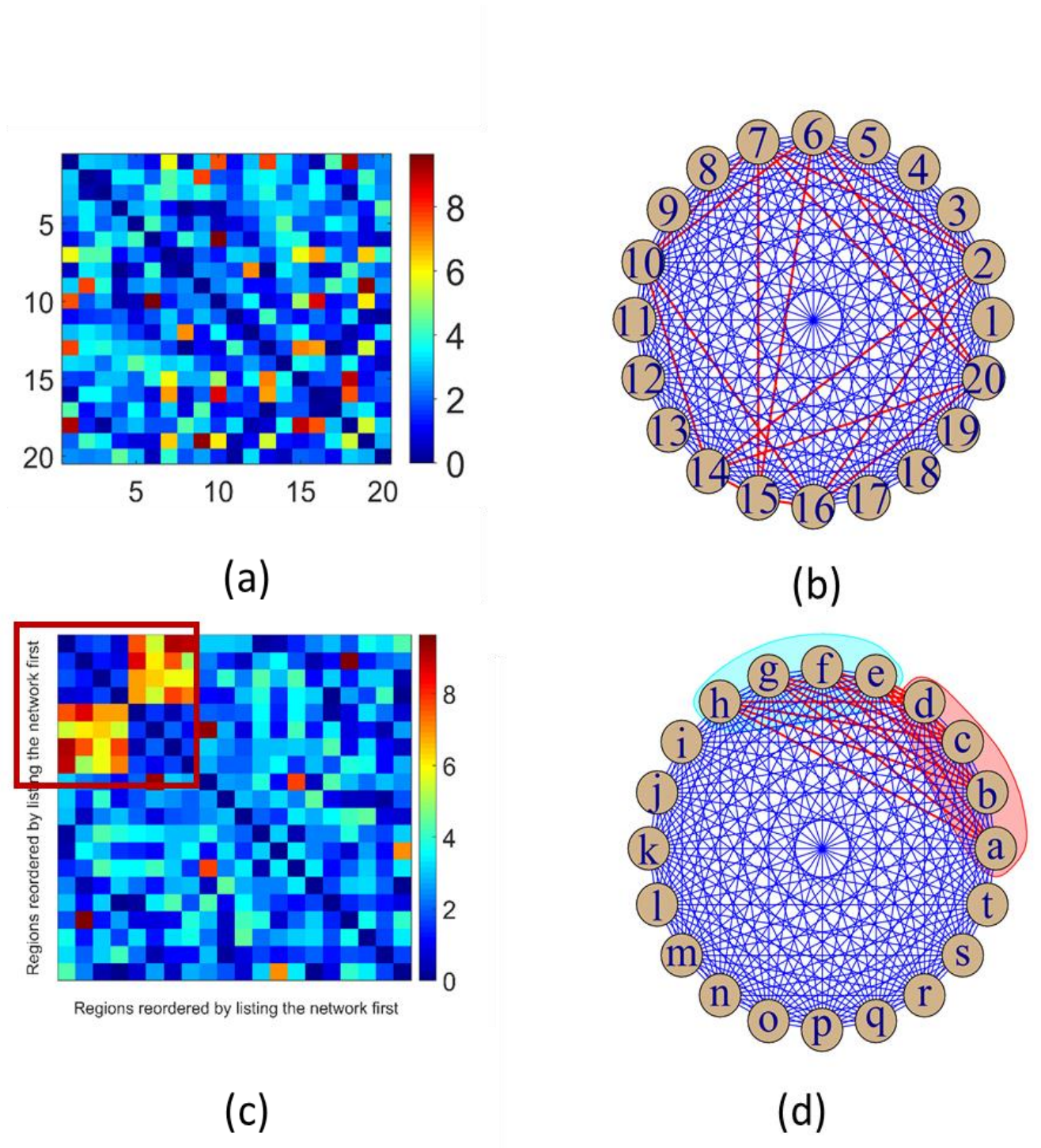

**Figure 1:** (a) A heatmap demonstrates group-wise connectivity statistical testing results in the form of -log(p); hotter points represent edges with more significant group differences. (b) A simple graph consists of the 20 nodes and 190 edges, where blue edges represent no (true) patient - control differences on connections between nodes, and red for true differences on connections between groups. (c) The reordered heatmap (the new node order is determined by the NOS) explicitly displays a disease-related subnetwork in the left-top corner, which shows a bipartite topological strucuture. (d) The reordered simple graph of (b) highlights the two disjoint sets of nodes of the bipartite subgraph in shades; the node order in (d) is the same as in (c). Note that (b) and (d) are isomorphic.

In comparison, NOS related procedures[13,14] can extract a latent differentially expressed connectomic subnetwork with a bipartite graph topology, where the term “subnetwork” denotes a brain connectomic subgraph with an organized graph topological structure that is statistically significantly different between groups or significantly contribute to a behavioral or disease feature. A bipartite graph is a graph whose nodes can be divided into two disjoint sets and an edge only exists when it connects two nodes, each from a disjoint set[14]. Figure 1c demonstrates the detected differentially expressed subnetwork in a red square. It represents a brain connectomic *circuitry* that is systematically different between patients and controls. Note that in Figure 1c we reorder the 20 ROIs by listing the nodes/ROIs in the detected bipartite subgraph first. The automatically detected bipartite graph topology is otherwise hidden from traditional individual edge based analysis, may lead to interesting findings. For example, the bipartite structure (Fig 1) may suggest that the within community connections are more reliable and well-wired for both patients and healthy controls, whereas the inter-community connections are more vulnerable and more likely to show group differences that could be related to disease related abnormal behaviors and brain functions[14].

The NOS method aims to extract and test latent, well-organized and disease related connectomic subnetworks from the population level whole brain connectomic data. The network level statistical inferential procedure is based on the graph combinatorics. We summarize the subnetwork detection and test framework (full mathematics are in the Methods section). NOS application is demonstrated using an rs-fMRI study for 70 schizophrenia patients and 70 matched controls (dataset 1 or **D1**), followed by a replicate data set of 30 patients and 30 controls (**D2**). The network-level findings are highly reproducible across the two data sets.

## Results

### Subnetwork detection algorithm and statistical inference overview

A basic clinical experiment is usually to detect the disrupted brain connectomic subnetworks by a given neuropsychiatric condition. We define the whole brain connectome by using a weighted complete graph $G=\{V,E\}$ with $n$ nodes and $n(n-1)/2$ edges where n is the number of brain areas. The weight of an edge in $E$ for an individual subject represents the connectivity metric between the two corresponding nodes. Let $G_k=\{V_k,E_k\}\subset G$ be a disrupted subnetwork that a high proportion of edges in $E_k$ are differentially expressed between the clinical groups, and we assume that $G_k$ demonstrates a well-organized graph topological structure (the structures are not limited to communities/cliques and could be k-partite, rich-club, and others). We propose a novel objective function to automatically identify $G_k$ (by relisting the order of nodes and regrouping the edges) such that most differentially expressed edges are included in the well-organized subnetworks and each subnetwork contains a high proportion of abnormal edges. The objective function consists of a penalty term of the (edge) size of the subnetwork $|E_k|$ to ensure that the extracted subnetworks are parsimonious, and the parsimonious property can effectively reduce false positive findings and improve the reproducibility. The constrained optimization results yield disease-related subnetwork objects that include three classes: nodes, differentially expressed edges, and the well-organized graph topological structure.

Next, new statistical inferential procedures are developed to formally test statistical significance of the detected subnetworks. Unlike conventional statistical methods that test on many of the individual metric (e.g. individual edge, an descriptive graph metrics, and/or a 'summed' component) followed by correction of multiple comparisons, we test the subnetworks as objects by jointly considering (i) the statistical significance of connectivity metrics on each edge across clinical groups and (ii) the distribution of the differentially expressed edges in the graph space from the graph combinatorics perspective. A true disease-related connectomic subnetwork is assumed to exhibit a high proportion of edges that are different between patients and controls. The chance of a single edge to be false positive or false negative is likely high; but the chance of a group of edges constrained by a well-organized graph topological structure being false positive or false negative is very rare. This statistical inferential procedure is better suited for disease-related network analysis compared to existing network analysis methods such as FWER, FDR, and NBS because the automatically detected topological structure allows edges borrow strengths from each other. Therefore, the dependence between edges is utilized to improve statistical inference via their latent graph topology though the covariance matrix between all edges is not explicitly estimated which is very challenging. The formal hypothesis testing is performed at the network-object level, which includes two testing strategies: 1) group label permutation 2) graph edge permutation. The new statistical inferential procedures can adjust for selection bias and multiplicity of subnetworks. The full mathematical description is in Methods and additional information in the Supplementary Information (SI) section.

**A clinical application of NOS**

The new methods are applied to an rs-fMRI data set of 70 patients with schizophrenia and 70 matched healthy controls, and the results are validated for replicability by using an independent data set of 30 cases and 30 controls collected at the same medical center but later in time. Imaging data processing procedures are described in SI. We define the nodes of the connectome graph $G$ by using 90 automated anatomical labeling (AAL) regions[16]. Time courses of all voxels within a region are pre-processed as region-wise signals, followed by calculating 4005 Pearson correlation coefficients between the time courses of the 90 AAL regions which are performed by Fisher's $Z$ transformation and empirical Bayes normalization[17] to obtain connectivity matrices for all subjects. Wilcoxon rank sum tests are used on the normalized correlation coefficients for all edges and the resulting 4005 p-values were stored.

We first examine whether any individual edge in $G$ is significant by applying multiple testing adjustment. The false discovery rate (FDR[18]) is used, and none of 4005 edges is found significant by using the threshold $q=0.2$. The network based statistics (NBS) also yields no differentially expressed structure by using various thresholds[11]. This is likely due to loss of power by multiple comparison adjustment without utilizing the topological and dependency structures.

Next, we perform subnetwork level analyses by extracting and testing the latent differentially expressed connectome subnetworks as objects. Let matrix $\mathbf{W}$ be the whole brain graph edge-wise testing result matrix, where entry *i*, *j* is $W_{ij} = -\log(p_{ij})$ where *i* and *j* are two distinct AAL regions and $p_{ij}$ is the corresponding test p-value for the edge between *i* and *j*. $\mathbf{W}$ is demonstrated in Figure 2a. We apply the latent subnetwork detection algorithms and statistical tests described in the method section to determine the significantly differential connectome subnetworks between patients and controls. The testing results show that one subnetwork is significant ($p < 0.001$). The significant subnetwork ($\mathbf{R}_1$ to denote the subnetwork from **D1**) includes 15 nodes, 60 altered edges, and a well-organized topological structure. The detected topology is a k-partite structure with $k=6$, that is $K_{1,1,1,1,1,10}$. Moreover, the rich-club coefficient of $G_1$ is very high (0.83)[19]. The first 5 nodes are rich-club nodes, and most of the 10 edges between these 5 nodes are differentially expressed between schizophrenia patients and healthy controls. The first five nodes include left insula cortex, the cingulate cortex, left Rolandic operculum, and bilateral paracentral lobules. The remaining 10 peripheral nodes include inferior temporal lobes, superior temporal lobes, Heschl gyrus, right insula cortex, and right Rolandic operculum. (Figure 1c) (full region names are listed in SI Table 1). This altered subnetwork is composed of approximately the fronto-parietal network and the cingulo-opercular network, which have been frequently associated with abnormalities in schizophrenia during attention, working memory, and executive control functional imaging studies[6,20]. Note that among the 60 differentially expressed edges, 59 edges show decreased or equivalent connections in patients with schizophrenia and only one shows hyper-connection. This may align with findings suggesting that schizophrenia is a 'dysconnectivity' disorder with primarily reduced functional connectivity across brain regions [20-22].

### Findings in a replicate data set

The same pre-processing steps and connectivity metric calculation are applied in another 30 patients and 30 healthy controls. We first compare D1 and D2 on the traditional edge-wise and network level statistical inferences on the 4005 edges, followed by comparisons of results using NOS.

*Comparing traditional edge-wise findings between **D1** and **D2***

As in D1, none of 4005 edges is found significant by using the threshold $q=0.2$ by FDR. Wilcoxon rank sum tests are then used to obtain the edge-wise testing p-values and matrix $\mathbf{W}^2$ using an arbitrary p<0.005 for both datasets. Patient-control differences are identified in 40 edges in **D1** and 52 edges in **D2**. However, only 6 of the 40 edges in **D1** are overlapped with the 52 edges in **D2**. Therefore, around 7% of the findings of analyzing **D1** and **D2** independently agree with each other. The cutoff p-value of 0.005 is used only for the purpose to evaluate the agreement of differentially expressed edges in **D1** and **D2.** The hazard of relying on node only or edge only information in fMRI studies are recently coming to its head with discovery that popularly used software to perform multiple comparison correction maybe under incorrect assumptions and produce unacceptably low replicability[38].

*Comparing edges only within the subnetwork between **D1** and **D2***

We next examine whether edges in the detected subnetwork of **D1** show differential expressions in **D2.** The region list of $\mathbf{W}^2$ is reordered based entirely on the detected significant subnetwork in **D1**, placing the ROIs of subnetwork $\mathbf{R}_1$ at the left-top corner (Fig 2c). We note that most edges (83%) in $\mathbf{R}_1$ also show group-wise difference (smaller p-values) in **D2:** among the 60 edges in $\mathbf{R}_1$, 59 edges show reduced or equivalent connectivity by comparing the median of connectivity metrics of patients to healthy controls in **D1**, while 58 edges in **D2** also demonstrate weaker or equivalent connectivity in schizophrenia. One of the two hyper-connected edges in **D2** is the one hyper-connected edge in **D1**. Therefore, both the signs and p-values of the edges agree with each other in two independent patient-control samples within the detected subnetwork $\mathbf{R}_1$ (by **D1**).

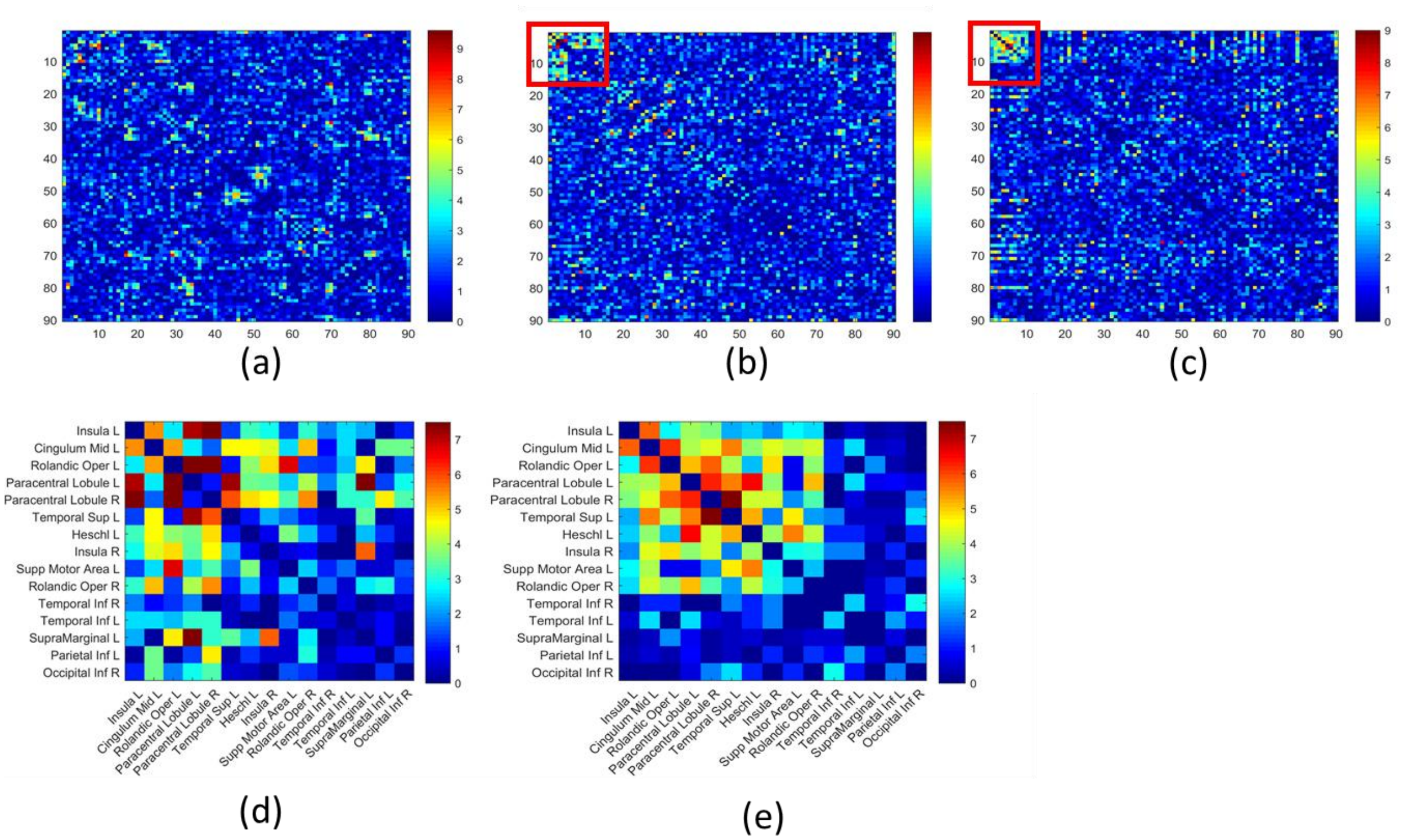

Figure 1: Application to clinical data and replication. (a) A heatmap of –log(p) of the first dataset (D1). A hotter pixel indicates more differential edge between cases and controls. There is no apparent pattern of these edges. (b) We then perform subnetwork detection and testing the algorithms, and find a significant subnetwork with the rich-club topological structure, where we reorder the nodes by listing the 15 nodes in the detected subnetwork (red square, which is magnified in d); (c) A heatmap of –log(p) of the second dataset (D2) where the order of nodes is the same of (b), and we find that most edges identified in the subnetwork of D2 are hot edges of subnetwork of D1 (red square, which is magnified in e).

*Comparing the full subnetworks detected from **D1** alone and **D2** alone*

The analysis above is started with **D1.** We now perform NOS on **D2** alone and compare it with $\mathbf{R}_1$. A subnetwork ($\mathbf{R}_2$) of 21 nodes in a clique structure is identified. Interestingly, we note that $\mathbf{R}_1 \subset \mathbf{R}_2$. Unlike edge-wise analysis that yields less than 7% of replications, the NOS method identifies the altered edges in the subnetwork from **D1** that can be completely rediscovered when analyzing **D2** independently, though the subnetwork in **D2** is larger (a clique of 21 nodes). This is likely because **D2** has more small p-value edges resulting in a larger network to be detected, as shown in Venn diagrams (Fig3). The positive agreement is used to compare reproducibility of features between **D1** and **D2** (statistical methods in SI) using NOS vs. individual edge based statistics. The network approach is significantly better than individual edge based method ($p<0.001$). In summary, by utilizing an independent replication data set collected posteriorly we can conclude that the findings identified by the NOS approach are more reproducible.

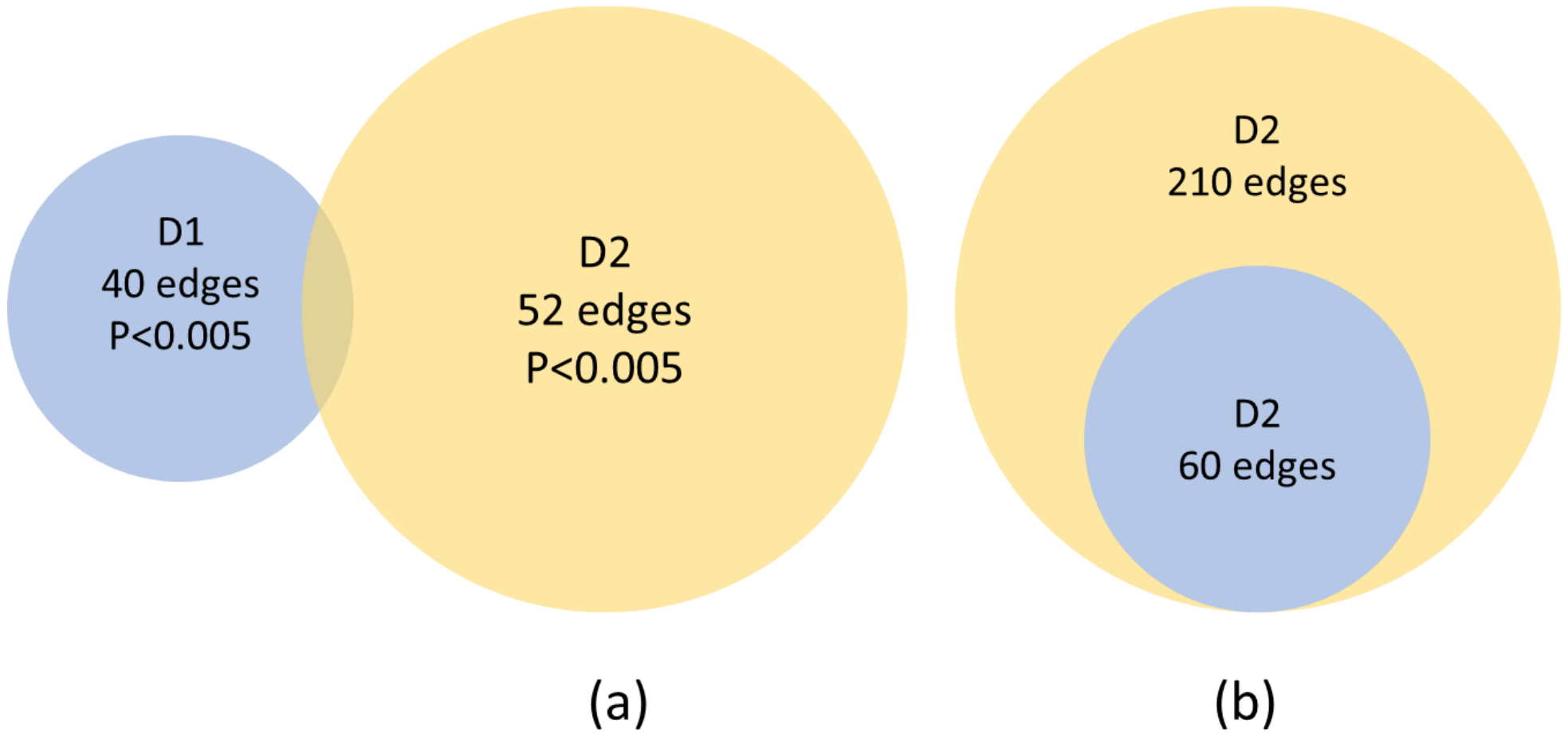

Figure 2: Comparing the subnetworks detected from D1 alone and D2 alone. (a): The overlapping edges by using traditional edge-wise inference. (b): the overlapping edges of two subnetworks detected by D1 alone and D2 alone using NOS methods.

## Discussion

In this paper, we propose NOS methods to solve a long-term challenge for brain connectomics analysis for discovering hidden, disease-related subnetworks. NOS makes several innovative contributions. Firstly, NOS introduces a new statistical framework to identify differentially expressed 'subnetworks' (instead of individual edges or univariate graph descriptive metrics) from population level whole-brain connectomic data, where each identified subnetwork is treated as an object. The object is defined (or constrained) by three components, i.e., nodes, edges, and latent graph topological structures, which consolidate the localized signals of nodes to individual edges to global graph topological information. Secondly, the new objective function implements automatic detection of latent differentially expressed connectome subgraph objects (rather than pre-defined) for the whole brain connectivity analysis. The shrinkage penalty term is novel and well-suited to extract information concentrated subgraphs parsimoniously from a weighted complete graph. The graph topological structures of extracted latent altered subgraphs are important for discovery of novel underlying circuit-level neuropathology. Last, we develop two formal statistical inferential procedures to rigorously test the differentially expressed subnetworks by leveraging graph combinatorics and information entropy, and the tests adjust the multiplicity and pre-selection bias.

The subnetwork findings of the schizophrenia rs-fMRI study reveal the latent disrupted functional connectome subnetwork that includes parts of the fronto-parietal network and cingulo-opercular networks. An independent data set collected posteriorly verifies the replicability of the subnetwork although the detailed graph topological structure differs in part. The results show that similar altered subnetworks are identified in both **D1** and **D2** by NOS with high proportion of overlapped edge-wise findings; in comparison, traditional edge-wise and network level inferences produce only a small overlap on edge-wise findings in **D1** and **D2**. The network-object statistics may provide a new pathway to conduct reproducible research using biomedical big omics data. Although we utilize functional connectivity from rs-fMRI data in this study, our methods are applicable to other brain connectomics data because this method only requires the input data as a set of connectivity matrices. Although AAL atlas and correlation coefficients as connectivity metrics are utilized for our data example, users are free to choose other human brain atlases and connectivity metrics based on the characteristics of their own data sets when applying

NOS[2].

NOS also demonstrates superior statistical properties. False negative and false positive discovery rates of group differences on edges are decreased by allowing edges borrowing strengths from each other within the same subnetwork. Both simulation and replication data sets verify this claim. From the graph combinatorics point of view, false positive/negative finding is likely when individual edges are compared, whereas the false positive/negative rate of a well-organized subnetwork should be quite small because the probability to observe such organized graph topology is rare. Although the information of each individual feature could be noisy from the biomedical ‘omics’ big data due to confounds from various sources, the new network approach provides an alternative solution through inter-linking a set of individual features and combining the information collectively by leveraging the subnetwork topology. Therefore, NOS appears empirically reliable and may improve the reproducibility of high-throughput omics study. In summary, this approach should open a new avenue for object oriented statistical analysis of the high-throughput, richly correlated connectomic and likely other omics data.

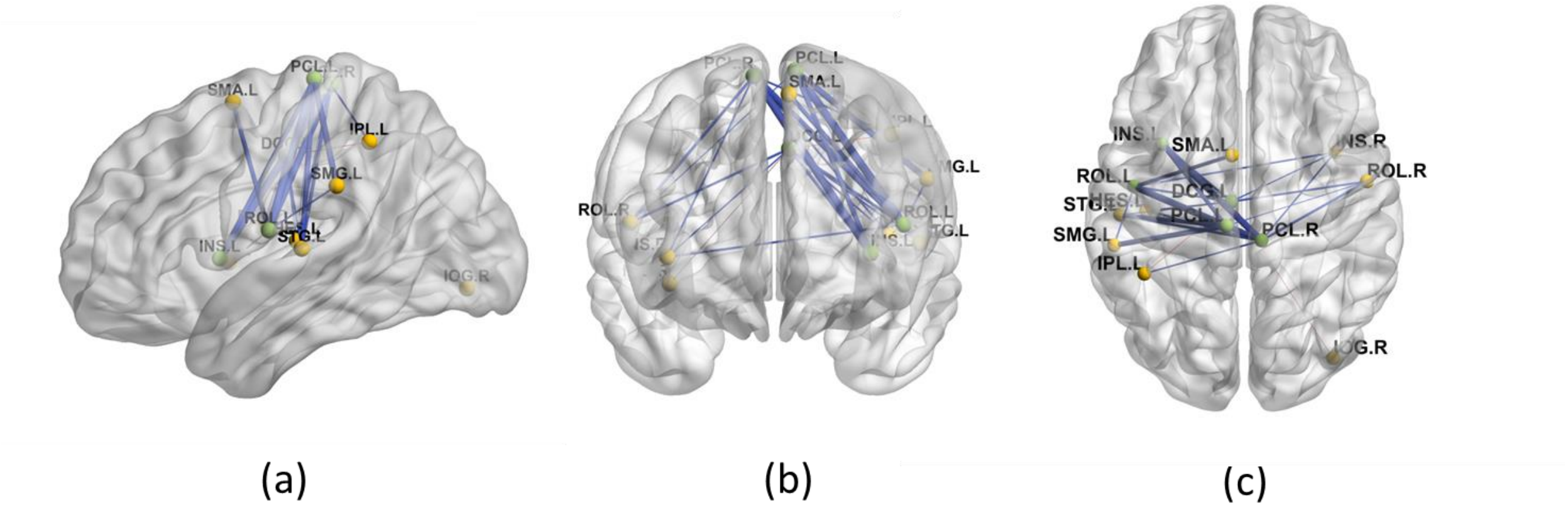

Figure 3. (a)-(c) display one snapshot of the edges in the subnetwork of D1 using 3D figures: the green

nodes are the five rich club-nodes and yellow nodes are the 10 peripheral nodes; we show the edges with p-value $< 0.005$ and the width of an edge reflects the statistical significance. Blue edges indicate the connections of normal controls are greater than patients, and red edges indicate the opposite. Full 3-D displays of results from D1 and D2 are in SI.

# Methods

### NOS Background

We introduce the NOS methods under the context of functional connectivity analysis, though the methods are applicable for connectivity analysis in other imaging and non-imaging modalities. Brain regions are denoted as nodes (vertices) and the connections between them are considered as edges [7,11,24]. The connectivity is often measured by a scalar metric, for instance, one commonly used connectivity metric is the Pearson correlation coefficient of the fMRI time series between a pair of brain areas.

Formally, let a symmetric matrix $\mathbf{M}^s$ denote the connectivity metrics of the whole brain connectome for subject $s$ ($s = 1, \cdots, S$). In a group level connectivity study of $S$ subjects, $\mathbb{M} = \{\mathbf{M}^1, \cdots, \mathbf{M}^S\}$ are connectivity matrices for all subjects. Each off-diagonal element $z_{i,j}^s = \mathbf{M}^s(i, j)$ is a transformed connectivity metric between areas $i$ and $j$, for example, Fisher's Z transformation for Pearson correlation coefficients. We use a complete graph $G = \{V, E\}$ to denote the whole brain connectomics. $V$ is the set of nodes (brain areas) and $E$ is the edge set (connections between brain areas), and all subjects share the same $V$ and $E$. But, the connection expression levels of edges in $E$ vary across subjects as in $\mathbf{M}^s$. Let vector $\mathbf{Z}^s = \underbrace{\{z_{1,2}^s, z_{1,3}^s, \cdots, z_{1,n}^s, z_{2,3}^s, \cdots, z_{2,n}^s, \cdots, z_{n-1,n}^s\}}_{n(n-1)/2}$ represent connectivity expressions for all edges in $G$ ($n = |V|$ is the number of brain areas), and it follows a (say normal) distribution that $\mathbf{Z}^s \sim \mathrm{MVN}(\mathbf{X}_s^T \boldsymbol{\beta}, \boldsymbol{\Sigma}_{\mathcal{G}})$. $\mathbf{X}_s$ is a vector of $q \times 1$ covariates (e.g. clinical and demographic variables) and $\boldsymbol{\beta}$ is a $q \times n(n-1)/2$ matrix for the effects of these covariates across all edges. Clearly, $\boldsymbol{\Sigma}_{\mathcal{G}}$ influences the statistical inferences of individual edges, and without properly accounting for $\boldsymbol{\Sigma}_{\mathcal{G}}$ the inferences on individual edges are subject to substantial false positive and false negative findings. Yet, the estimation of $\boldsymbol{\Sigma}_{\mathcal{G}}$ is complicated because i) it contains a massive number of parameters: for n=300 brain areas the number of parameters in $\boldsymbol{\Sigma}_{\mathcal{G}}$ is $[n(n-1)/2] \times [n(n-1)/2-1]/2 \doteq 1.2 \times 10^9$; ii) massive parameters are constrained by the spatial geometry (a pair of edges involve 4 nodes with 3D coordinates). Therefore, we develop NOS to account for the dependence structure by imposing graph topological constraints.

We refer to a subnetwork $G_k$ as a subgraph of the whole brain connectome graph $G$ with any topological structure ($G_k \subset G$). $G_k$ is not necessarily a community (i.e. clique and induced subgraph of $G$) and it may be a k-partite or a rich-club subgraph. We denote a latent disrupted subnetwork $G_k = \{V_k, E_k\}$ as a subgraph $G_k \subset G$, where $k = 1, \cdots, K$ and $1 < K < |V|$. Inside the subgraph $G_k$, edges show altered connectivity expressions across clinical subgroups, for instance, $E(z_{i,j}^{control}) \neq E(z_{i,j}^{case}) \mid e_{i,j} \in E_k$. Let the vector $\mathbf{M}_k = \{z_{i,j}, \forall e_{i,j} \in E_k\}$ be the differentially expressed connectivity metrics. We characterize the differentially expressed subgraph $G_k$ by a network-object $\mathbf{R}_k = \{V_k, \mathbf{M}_k, \mathcal{T}_k(V_k, E_k)\}$ including three components: 1) the differentially expressed connections $\mathbf{M}_k$ for the edge subset $E_k$; 2) $V_k$ as the set of nodes connected by $E_k$; and 3) an organized graph topological structure. Using the example in the introduction section (Figure 1a - d), $G_k$ includes 8 nodes and 16 edges, and $\mathcal{T}_k(V_k, E_k)$ is a bipartite graph topology that the set of nodes are split into two sets $V_k = V_k^1 \cup V_k^2$ such that $e_{i,j} \in E_k$ if and only if $i \in V_k^1$ and $j \in V_k^1$. In general, $\mathbf{R}_k$ is a subgraph based strongly non-Euclidean object oriented statistic but the statistical inference for such a subgraph may be challenging[25]. Instead, we construct the NOS methods by integrating statistical inferences and graph combinatorics, which effectively allows researchers to obtain statistical significance values on network-based analysis.

## Subnetwork extraction

We propose a fundamentally new heuristic method to extract latent and systematically disrupted subnetworks from the whole brain connectome. The method first assigns a weight $w_{i,j}$ for each edge in the connectome $e_{i,j} \in E$ to denote the difference between clinical groups, for example, $w_{i,j} = -\log(p_{i,j})$ where $p_{i,j}$ is the corresponding two sample test p-value or other more general linear regression analysis p-value. Let $\mathbf{W}_{n \times n}$ be the symmetric weight matrix. We define $U = \cup_{k=1}^{K} \mathrm{G}_k$ as a union set of subgraphs ($1 \le K \le |V|$ and $G_k = \{V_k, E_k\}$) with the following conditions:

1. $\cup_{k=1}^{K} V_k = V$;
2. $\cup_{k=1}^{K} E_k \subseteq E$, when $K = 1, \cup_{k=1}^{K} E_k = E$ whereas given $K = |V|, \cup_{k=1}^{K} E_k = \varnothing$.

To extract subnetworks is equivalent to estimate the union set of subgraphs $\cup_{k=1}^{K} \mathrm{G}_k$ and the number $K$. Our goal is to select the organized subgraphs that cover most differentially expressed edges while keeping high proportions of differentially expressed edges in the subgraphs. We implement this goal by optimizing an objective function that maximizes the sum of weights covered by $U = \cup_{k=1}^{K} \mathrm{G}_k$ with the **constraint** to minimize the size of the union set, that is, $|U| = |\cup_{k=1}^{K} \mathrm{G}_k| = \sum_{k=1}^{K} |\mathrm{E}_k|$ :

$$\underset{\cup_{k=1}^{K} G_k, K}{\arg\max} \sum_{k=1}^{K} \exp\{\log(\sum w_{i,j} \mid (e_{i,j} \in G_k)) - \lambda_0 \log(|\mathrm{E}_k|)\}, \quad (1)$$

where $\lambda_0$ is a tuning parameter between 0 and 1 (see SI). Note that the subgraphs in $U = \cup_{k=1}^{K} \mathrm{G}_k$ are limited to those with organized topological structures, for instance, k-partite, clique, and rich-club subgraphs. NOS imposes the rule of parsimony: for any given constraint value $\lambda_0$ , only a small subset of edges $\cup_{k=1}^{K} E_k \subset E$ can be included in the union set of subgraphs to cover the differentially expressed edges. Therefore, we consider the general NOS subgraph extraction objective function as a shrinkage estimator of the network object (shrinking the size of the network-object to zero). We have recently developed subgraph detection algorithms including parsimonious connectivity network detection (Pard,[13] for the clique structure) and k-partite graph detection (KPGD,[14] for the multipartite structure). We next determine the statistical significance of the extracted subgraphs.

## Testing Hypotheses

One of the overarching goals of brain connectome analysis is to answer whether the connectivity patterns are differentially expressed between different clinical groups. Here, NOS for hypothesis-testing is illustrated in two-group comparisons. However, similar hypothesis testing steps using NOS can be applied to other connectomics questions, for example, whether connectivity patterns are related to specific clinical and biological features or are changed over time or by specific treatments.

*Testing individual edges vs. subnetworks*

Most conventional statistical methods perform multivariate edge-wise statistical inferences on brain connectome analysis, for instance, the family-wise error rate (FWER), the false discovery rate (FDR), the network based statistics (NBS), and various versions of the least absolute shrinkage and selection operator (Lasso). Generally, the selection of features is solely dependent on the association between individual features and outcome variable (e.g. the p-values and test statistics) in these methods without fully accounting for the correlation structure between all features. However, in biomedical big data analysis the true signals are often overwhelmed by the false positive and false negative noises, which may cause low reproducibility of findings and limited clinical applications. When a connectome subnetwork $G_k$ is truly altered in cases compared with controls, performing edge-wise inference may false negatively exclude many edges in $G_k$ as they cannot meet the stringent (adjusted) threshold, and false positively include non-differential edges. Therefore, we may miss the opportunity to identify $G_k$ in the study.

One alternative way is to test on the network-object $\mathbf{R}_k$ directly based on the extracted $G_k$ by examining whether $\mathbf{R}_k$ is differentially expressed between groups. The formal statistical testing hypotheses are:

$H_0$ : There is no differentially expressed connectome subnetwork when comparing the connectivity patterns across clinical populations, which is equivalent to:

C1: there are no differentially expressed edges; or C2: there are differentially expressed edges, but they are randomly distributed in the whole graph $G$.

$H_1$: There is (are) differentially expressed connectome subnetwork(s) in $G$ that is equivalent to:

C1: there are differentially expressed edges; and

C2: the differentially expressed edges are NOT randomly distributed in $G$, and the connectivity metrics $\mathbf{M}_k$ within the detected topology $\mathcal{T}_k(V_k, E_k)$ are more likely to be differentially expressed across clinical groups.

Statistical procedures have been developed to conduct the omnibus test for C1 e.g. adaptive sum of powered score (aSPU) test[26]. If we fail to reject C1, then we fail to reject $H_0$. Given C1 is rejected, we test C2 by examining the 'spatial' distribution of altered edges in the space of the whole brain connectome graph $G$: whether the altered edges are clustered in $G_k$. Spatial statistical methods have been used to test clustering and clusters of (incidence) points on the geographical maps, for instance, in cancer surveillance to identify areas of elevated risk and to investigate hypotheses about cancer etiology[27-30]. The popular tools such as Kulldorff's spatial scan statistic and K functions are developed based on the spatial point process. However, the features of interest in connectomic data analysis are connectivity edges rather than incidence points, for which the cluster detection methods in scan statistics are not directly applicable. The NOS subnetwork extraction methods provide the window of a cluster by $\mathcal{T}_k(V_k, E_k\}$. Therefore, the NOS inferential procedure is built based on graph combinatorics.

## Monte Carlo Tests

We develop two strategies to examine the hypotheses above by taking the graph combinatorics into consideration: the first strategy tests C1 and C2 simultaneously and the second strategy tests them sequentially.

## Group label permutation

Strategy one is group label permutation (GLP). Group label shuffling technique is widely used to extract altered brain activation[29] and connectivity features[11]. The group label permutation randomly assigns each subject's clinical group label for each iteration. The machinery of group label shuffling is that with shuffled group labels the brain connectivity levels are expected to show no difference between (shuffled) clinical groups ( $H_0$: C1); and even though some edges are false-positively significant they are not likely to be distributed in an organized pattern as the original data set (C2). Therefore, in each iteration the GLP simulates a data set under the null hypothesis that follows both C1 and C2.

*Test statistic*

For each group label permuted data set, we perform group level statistical analysis by applying statistical tests and the Pard and KPGD algorithms, identify the potential subnetworks, and then store the maximum test statistic of all detected subnetworks by (1). The test statistic is used to summarize difference between 'groups' regarding all edges collectively in the detected brain connectivity subnetwork $\mathbf{R}_k^m = \{V_k^m, \mathbf{M}_k^m, \mathcal{T}_k^m(V_k, E_k)\}$ , where $m = 1, \cdots, \mathcal{M}$ and $\mathcal{M}$ is the total number of simulations. A natural choice of the test statistic is motivated by the scan statistic as:

$$T_{max,scan}^m = \max_{k=1,\cdots,K^m} \left(\frac{N_{1,in}}{N_{in}}\right)^{N_{1,in}} \left(\frac{N_{1,out}}{N_{out}}\right)^{N_{1,out}} I\left(\frac{N_{1,in}}{N_{in}} > \frac{N_{1,out}}{N_{out}}\right),$$

where $N_{1,in}$ is the number of edges inside a subnetwork with univariate testing p-value $< p_0$ ( $p_0$

is a pre-defined threshold) $N_{1,in} = \sum_{e_{i,j} \in E_k} I[\hat{p}_{i,j} < p_0]$ and $N_{in}$ is the total number of edges inside the network. $N_{1,\text{out}} = \sum_{e_{i,j} \notin E_k} I[\hat{p}_{i,j} < p_0]$ and $N_{out}$ are for edges outside of the subnetwork correspondingly. Max indicates we store the test statistic with the maximum value among all subnetworks in the $m$ th iteration.

However, $T^m_{max,scan}$ is subject to the arbitrary choice of $p_0$ and does not discriminate the difference for edges with testing p-values $< p_0$ (e.g. no difference regardless p value is 0.04 or 0.000001 when using $p_0 = 0.05$). Fisher's combined probability test statistic seems to avoid these two disadvantages [31;32], as it summarizes all edges in a subnetwork $\mathbf{R}^m_k$ . For the $m$ th iteration, we first calculate the $\chi^2$ statistic for each detected subnetwork by using $x^m_k = -2 \sum_{i,j \in \mathbf{R}^m_k} \log p^m_{ij}$ . Since the detected subnetworks may have various sizes regarding number of nodes and edges, it is difficult to compare the test statistic across these subnetworks. Instead, we utilize the corresponding probability: $Prob(\mathbf{R}^m_k) = 1 - F_{\chi^2,\, df=2|E^m_k|}(x^m_k)$ where $F$ is the cumulative probability function of $\chi^2$ distribution with degree of freedom equal to $2|E^m_k|$. A smaller $Prob(\mathbf{R}^m_k)$ represents more difference between clinical groups. The percentile of probability reflects the graph's two perspectives jointly: the collective difference of connectivity in the subnetwork and the organized distribution of differential edges. In practice, the p-value is very small when the number of edges $|E^m_k|$ is relatively large and the proportion of differential edges is high. We adopt the Chernoff bound of the $\chi^2$ cumulative distribution function to boost the computational speed and to assess the minor difference for large test statistics [33]. Thus, the test statistic is

$$T^{m}_{max,Fisher} = \max_{k \in K^m}\{-\log(Prob(\mathbf{R}^m_k))\} = \max_{k \in K^m}\{-|E^m_k|\log(\overline{x}^m_k \exp(1-\overline{x}^m_k))\} = \max_{k \in K^m}\{|E^m_k|(\overline{x}^m_k - 1 - \log(\overline{x}^m_k))\}$$

where $\overline{x}^m_k = -2\sum_{i,j \in G^m_k} \log p_{i,j} / (2|E^m_k|)$ is an empirical estimate of information entropy of the selected subgraph.

Then, we perform $\mathcal{M}$ times of permutations and for each iteration $m$ we store $T^{m}_{max,Fisher}$. If the observed test statistic $T^0_k$ for the detected subnetwork $k$ is among the highest $\alpha$ (e.g. 5%) percentile of distribution of all $T^{m}_{max,Fisher}$, we consider it is significant[11,27-29]. The robustness of the statistics is improved by allowing edges within a network-object to borrow strengths from each other under the guidance of well-defined topology in $\mathbf{R}_k$. Meanwhile, the detected topology in $\mathbf{R}_k$ also assists to control and lower false positive findings. The overall GLP algorithm is in SI Algorithm 1.

**Graph edge permutation**

Strategy two is to statistically examine the two conditions (C1 and C2) sequentially. In statistical literature, methods have been developed to perform the omnibus test to examine condition C1 (e.g. Zhang et. al. 2014). If the omnibus test is rejected such that the difference exists, we next examine condition C2 on whether the differentially expressed edges are clustered in the detected subnetworks. One Monte Carlo testing strategy is to permute the graph to generate Erdős and Renyi random graphs and test whether the altered edges are distributed randomly.

Similar to GLP, for the $m$ th iteration the GEP algorithm calculates the test statistics for all detected subnetworks and stores the maximum $T^{m}_{max,Fisher}$. If the observed subnetwork $G_k$ is a genuinely altered object, the test statistic $T_k^0$ is expected to be large. Thus, we reject the null hypothesis if $T_k^0$ is among the top 5% percentile based on the distribution of the maximum test statistics of edge-shuffled graphs. We summarize the detailed algorithm in SI Algorithm 2. Moreover, the GEP test can be generalized to compare refined topological pattern (e.g. K-partite subgraph) with clique for a subnetwork $G_k$ [14]. In general, the GLP algorithm is similar to the case-control scan statistic method, whereas the GEP algorithm is linked with the incidence/case only scan statistic method. We include the detailed descriptions regarding graph combinatorics, multiplicity and selection bias, algorithms and others in SI.

## Performance and sensitivity analysis

We first evaluate the performance on efficiency and false positive rates of NOS by using simulated data. For group level connectome data sets $\mathbb{M} = \{\mathbf{M}^1, \cdots, \mathbf{M}^S\}$ and the corresponding graph whole brain connectome $G$, we define a subnetwork $G_k \subset G$ where edges are differentially expressed between controls and cases. The transformed (e.g. Fisher's Z) connectivity metric of each edge is set to marginally follow a normal distribution with $\mu_0$ and $\sigma_0^2$ (for controls) and $\mu_1$ and $\sigma_1^2$ (for cases). For differentially expressed edges (i.e. $e_{i,j} \in E_k$ ), $\mu_0 = \mu_1 + \theta$ and otherwise $\mu_0 = \mu_1$. Also, we let $\sigma_0^2 = \sigma_1^2 = \sigma^2$. We set a compound symmetry covariance matrix for edges within the subnetwork (correlation around 0.3), and the rest of edges are independent. By shuffling the order of nodes in $G$, the altered connectivity subnetwork is latent in the data set. In the simulation, we let $|V| = 100$, $|V_k| = 20$, and $G_k$ to be a clique (and $|E_k| = 190$). Two sample sizes (60 and 120) are used to represent the commonly observed sample size from a single study.

Each setting is simulated with different $\theta, \sigma$ and number of subjects for each group for 100 times.

The subnetwork detection algorithm and GLP and GEP for statistical inference are applied to each simulated data set. The efficiency is evaluated by both subnetwork and edge level false negative discovery rates; their false positive finding numbers are also reported. Note that edge-wise power (efficiency) can be further calculated as $1-FN/190$. Our method is compared with other multiple testing methods including Benjamini–Hochberg false discovery rate control (FDR) and local false discovery rate control (*fdr*)[23]. The false positive (FP) findings (*number* of FP edges in mean and standard deviation across 100 repetitions) and the according false negative (FN) edges are shown in **Table 1**.

Table 1: Simulation results of the NOS methods and the comparisons

| number of subjects: 30 vs. 30 | | | | |
|---|---|---|---|---|
| | $\sigma$ | 0.5 | 1 | 2 |
| GLP | FP | 10.7(28.53) | 11.8(30.36) | 9.25(18.38) |
| | FN | 4.95(22.13) | 3.75(16.77) | 0(0) |
| | Network FP | 0(0) | 0(0) | 0(0) |
| | Network FN | 0(0) | 0(0) | 0(0) |
| GEP | FP | 7.15(15.25) | 8.25(18.31) | 16(36.13) |
| | FN | 0(0) | 2.7(12.07) | 11.3(42) |
| | Network FP | 0(0) | 0(0) | 0(0) |
| | Network FN | 0(0) | 0(0) | 0(0) |
| FDR | FP | 45.98(9.07) | 43.37(9.57) | 31.11(8.01) |
| | FN | 8.23(3.02) | 24.99(5.37) | 70.67(8.12) |
| *fdr* | FP | 1.01(1.05) | 0.74(1.05) | 0.11(0.4) |
| | FN | 57.26(12.56) | 101.58(16.11) | 175.95(14.52) |
| number of subjects: 60 vs. 60 | | | | |
| | $\sigma$ | 0.5 | 1 | 2 |
| GLP | FP | 2(6.16) | 2.05(9.17) | 8.1(13.86) |
| | FN | 0(0) | 0(0) | 0(0) |
| | Network FP | 0(0) | 0(0) | 0(0) |
| | Network FN | 0(0) | 0(0) | 0(0) |
| GEP | FP | 3.05(9.99) | 2.05(9.17) | 11.8(30.13) |
| | FN | 0(0) | 0(0) | 0(0) |
| | Network FP | 0(0) | 0(0) | 0(0) |
| | Network FN | 0(0) | 0(0) | 0(0) |
| FDR | FP | 54.55(8.35) | 51.6(11.77) | 47.85(7.53) |
| | FN | 0(0) | 0.25(0.44) | 5.85(2.39) |
| *fdr* | FP | 0.1(0.31) | 0.5(0.61) | 0.8(1.32) |
| | FN | 5.6(2. 54) | 17.55(5.4) | 56.8(7.63) |

*fdr*: local false discovery rate control; FDR: Benjamini–Hochberg false discovery rate control (FDR)

Both GLP and GEP show excellent performance on network level inference by identifying the latent and differentially expressed subnetwork with 0 FP and FN rates. Next, GLP and GEP methods (based on the selected subnetwork) are compared with FDR and local *fdr* at individual edge inference using $q = 0.2$ as cut-off for both FDR and *fdr*. The results show that generally FDR has higher FP but lower FN rates compared with *fdr* (i.e. *fdr* is more conservative). Importantly, GLP and GEP methods outperform FDR and *fdr* when jointly considering FP and FN rates, see **Table 1**. Finally, the GLP and GEP methods are compared with the NBS method, but no subnetwork is detected by NBS when tuning the cutoff parameter from 3 to 6 for all settings (not shown in Table 1). One possible reason of the NOS methods over-performing the others is that, again, NOS allows edges to borrow strength with each other within the detected subnetwork object $\mathbf{R}_k$ . The GLP method seems to be more robust to the noise level than GEP. In summary, the NOS inferential procedures demonstrate excellent performance for testing altered subnetwork and providing edge-wise inference.

On Type I error rate, we count the number of false positive significant subnetworks for the data sets with no differentially expressed connectome networks (e.g. $\theta = 0$ ). Based on simulation of 1000 iterations, the network level false positive rates of GLP is 1.2% and GEP is 2.9%. Therefore, the network level Type I error is well controlled and below the subnetwork claimed $\alpha$ level of 5%.

## Clinical Samples

**Testing sample**: We recruited 70 individuals with schizophrenia (age = 40.80 $\pm$ 13.63 years) and 70 control subjects (age = 41.79 $\pm$ 13.44 years) matched on age (t=0.62, p=0.54) and sex ratio ( $\chi^2$ =0, p=1). All participants provided written informed consent that had been approved by the University of Maryland Internal Review Board. All participants were evaluated using the Structured Clinical Interview for the DSM-IV diagnoses. We recruited medicated patients with an Axis I diagnosis of schizophrenia through the Maryland Psychiatric Research Center and neighboring mental-health clinics. We recruited control subjects, who did not have an Axis I psychiatric diagnosis, through media advertisements. Exclusion criteria included hypertension, hyperlipidemia, type 2 diabetes, heart disorders, and major neurological events, such as stroke or transient ischemic attack. Illicit substance and alcohol abuse and dependence were exclusion criteria.

**MRI acquisition and pre-processing**: Data were acquired using a 3-T Siemens Trio scanner equipped with a 32-channel head coil at the University of Maryland Center for Brain Imaging Research. A T1-weighted structural image (MP-RAGE: 1 mm isotropic voxels, 256 x 256 mm FOV, TR/TE/TI = 1900/3.45/900ms) was acquired for anatomical reference. Fifteen minutes of resting state functional imaging was collected on each subject. During the resting scans, subjects were given a simple instruction to rest and keep their eyes closed. Head motion was minimized using foam padding, foam molding, and tapes. Resting-state fMRI were acquired over 39 axial, interleaving slices using a gradient-echo EPI sequence (450 volumes, TE/TR = 27/2000 ms; flip angle = $90^o$; FOV = 220x220 mm; image matrix = 128x128; in-plane resolution 1.72x1.72mm. Following the previously published procedures[35, 36], data were preprocessed in AFNI[35] and MATLAB (The MathWorks, Inc., Natick, MA). More details of the preprocessing steps are in SI.

**Replication sample**: We recruited another 30 individuals with schizophrenia (age = 39.73 $\pm$ 13.79 years) and 30 control subjects (age = 39.73 $\pm$ 14.16 years) matched on age (t=0.27, p=0.78) and sex ratio ( $\chi^2$ =0.09, p=0.77), following the initial sample of 70/70. The recruitment procedures, inclusion and exclusion criteria, and imaging acquisition and preprocessing procedure were kept the same. None of the testing or replication samples were previously published.

## Supplement i: Details of Permutation test

### *Graph combinatorics and permutation*

The NOS inferential procedure is built based on graph combinatorics. For instance, suppose there are roughly 10% of all edges with the univariate test p-value $<$ 0.05. If these suprathreshold edges are distributed randomly ($G$ is an Erdős and Renyi random graph), within a graph topology structure $\mathcal{T}_k(V_k, E_k\}$ we expect to observe around $|E_k|\times 10\%$ suprathreshold edges. Based on graph combinatorics the probability for $\mathbf{R}_k$ containing 50% or more suprathreshold edges is close to zero. The probability is lower for a larger size of $|E_k|$. For example, within a clique subgraph of 5 nodes and 10 edges the probability to observe 50% or more ($\geq 5$) suprathreshold edges is 0.001, whereas in a clique subgraph of 10 nodes and 45 edges the probability of (suprathreshold edges $\geq 23$) is $6\times 10^{-10}$.

### *Graph edge permutation vs. graph node permutation*

There are two sets of a graph: a set of vertices and a set of edges as $G=\{V,E\}$. Correspondingly, there are two options of permutation: permuting nodes or edges. First, we consider the permutation of nodes as a reordering process $\pi$, which is an 'edge-preserving bijection'. If two nodes $a$ and $b$ are connected in graph $G$, then in the node-permuted graph $H=\pi(G)=\{\pi(V),F\}$, then $\pi(a)$ and $\pi(b)$ are connected: $E_{ab}=1 \Leftrightarrow F_{\pi(a)\pi(b)}=1$.

$G$ and H are isomorphic graphs $G \simeq H$ (Figs 0 and 0). The network detection algorithms such as Pard and KPGD algorithms are essentially heuristic guided node-permutation methods. The original graph is a mixture of a block diagonal graph or random graph or K-partite graph, yet the topological patterns are implicit. These algorithms reorder the nodes to uncover these latent topological patterns.

Contrastingly, the edge permutation in Algorithm 2 is different because it permutes the order of edges and is not 'edge-preserving bijection'. For example, two nodes $a$ and $b$ are connected in a graph $G$; the edge-permuted graph $L$ (e.g. by permutation $\mu$) that $L = \mu(G) = \{V, F = \mu(E)\}$ and in $L$, $a$ and $b$ are only connected with probability of $p_G$, where $p_G$ is number of connected edges in $G$ divided by $n \times n / 2$. Therefore, the above two events are independent: $\{E_{ab} = 1\} \perp \{F_{ab} = 1\}$. Hence, though there is an organized pattern in $G$, the edge permuted graph $L = \mu(G)$ becomes a random graph without any organized patterns. The connectivity testing p-value matrix after edge permutation represents a random graph where each edge has the identical probability such that $p_{i,j} < p_0$.

***Adjustment for multiple testing and pre-selection bias***

If $G_k$ has been known from prior knowledge before data analysis, the statistical inferences boil down to a multivariate testing problem regarding $\mathbf{M}_k$. However, although prior neurobiology and clinical knowledge offer a guide, they are typically incomplete knowledge. The full $G_k$ is hidden in most studies and cannot be adequately pre-defined, which implies a complicated multiple testing problem for potentially infinite number of choices of $G_k$ for a given $G$. For example, for a graph $G$ with 100 nodes the number of possible 'clique' subnetworks is around $1.27 \times 10^{30}$. Monte Carlo and permutation tests have been used to account for the multiplicity but replications of initial findings remain challenging [11,27-29]. In addition, these permutation testing procedures do not specify their sizes or locations before cluster detection in each simulation iteration and prone to selection bias (i.e. the so-called 'Texas Sharpshooter' problem[22]). Therefore, new Monte Carlo and permutation tests are needed in the graph space to conduct subnetwork level statistical inferences.

***Test statistics jointly accounting for subnetwork size and average in-network edges' statistical significance level***

When selecting and testing the differentially expressed subnetwork $G_k$ , we are facing the tradeoff of the size of network and in-network edges' average statistical significance level. If the permutation test statistic focuses solely on the in-network edges' average statistical significance level, then the most significant edge within a network of two nodes will achieve the highest significance level and no large subnetwork will be significant. Thus, we need to jointly consider the size of network and the testing significance levels of in-network edges. The proposed test statistic $T^m_{max,Fisher} = \max_{k \in K^m}\{|E^m_k|(\bar{x}^m_k - 1 - \log(\bar{x}^m_k))\}$ with $\bar{x}^m_k = -2\sum_{i,j \in G^m_k} \log p_{i,j} / (2|E^m_k|)$ is derived based on Fisher's combination test and Chernoff bound of $\chi^2$ cumulative distribution function, which provides a practical solution to this challenge. In the contour plot (Fig1S) below, we demonstrate how the test statistic is affected by the two factors: the size of subnetwork and average testing significance level in context of community subnetwork. For example, the test statistic of a clique subnetwork of size 20 (nodes) and average testing significance of -log(0.1) is equal to a subnetwork of size 8 (nodes) and average testing significance of -log(0.0001). Therefore, the test statistic jointly evaluates the information quality and quantity of the selected network.

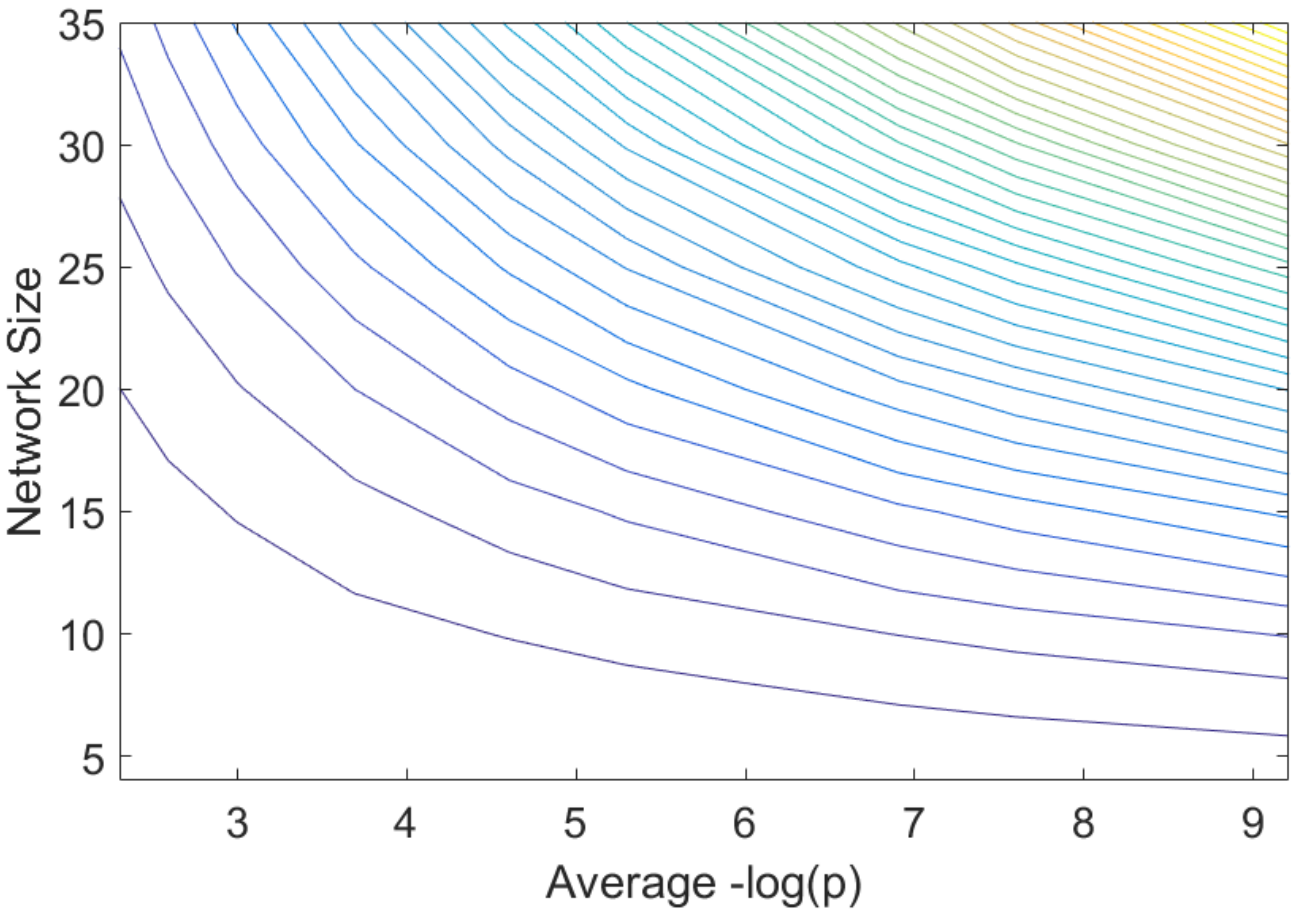

Figure 4S: Test statistic contour plots: network size (of nodes) and average -log(p)

We list the detailed algorithms of group label permutation and edge label permutation as follows:

---

**Algorithm 1** Group Label Permutation (GLP) Test

---

1: **procedure** GLP –Algorithm

2: **for** each permutation iteration $m = 1 : \mathcal{M}$ **do**

3: Shuffle group label for each subject;

4: Perform statistical tests on each edge based on shuffled grouped labels;

5: Apply the subnetwork detection algorithms[34] on $\mathbf{W}^m$;

6: Calculate the test statistic $T^m_{max,Fisher} = \max_{k \in K^m} \{ | E^m_k | (\bar{x}^m_k - 1 - \log(\bar{x}^m_k)) \}$ for $k = 1, \cdots, K^m$.

8: **end for**

9: Using the original group labels of all subjects, we apply the subnetwork detection algorithm and $T^0_k = | \mathbf{E}^0_k | (\bar{x}^0_k - 1 - \log(\bar{x}^0_k))$ for $k^0 = 1, \cdots, K^0$.

10: If $T^0_k$ is greater than 5th percentile of $T^m_{max,Fisher}$, we reject the Null Hypothesis, and thus $G^0_k$ is a differentially expressed subnetwork.

11: **end procedure**

---

**Algorithm 2** Group Edge Permutation (GEP) Test

1: **procedure** GEP –Algorithm

2: We perform SPU test ([26]) for the whole graph;

3: **If** the SPU test is not significant, **then** no SUBNETWORK is found;

4: **Else** Perform graph edge permutation (GEP) Test:

5 **for** each permutation iteration $m = 1 : \mathcal{M}$ **do**

6: List the all edges as a vector, $E = \{e_{1,2}, \cdots, e_{|V|-1,|V|}\}$;

7: Shuffle the order of edges in $\mathrm{E}$ and $-\log(p_{ij})$ accordingly, and obtain $\mathbf{W}^m$;

8: Apply the subnetwork detection algorithms (e.g. Pard) and obtain $K^m$ subnetworks;

9: Calculate and store the test statistic $T^m_{max,Fisher}$;

10: **end for**

11: Apply the subnetwork detection algorithms for the original data set and calculate $T^0_k$ for for each detected subnetwork $k^0 = 1, \cdots, K^0$

12: If $T^0_k$ is greater than 5$^{th}$ percentile of $T^m_{max,Fisher}$, we reject the Null Hypothesis, and thus $G^0_k$ is an SUBNETWORK with statistical significance.

13: **end procedure**

## Supplement ii: preprocessing of the fMRI data

Volumes were slice-timing aligned and motion corrected to the base volume that minimally deviated from other volumes using an AFNI built-in algorithm. After linear detrending of the time course of each voxel, volumes were spatially normalized and resampled to Talairach space at 3x3x3 $mm^3$, spatially smoothed (FWHM 6 mm), and temporally low-pass filtered (fcutoff = 0.1 Hz). For functional connectivity analyses, the six rigid head-motion parameter time courses and the average time course in white matter were treated as nuisance covariates[37]. A white matter mask was generated by segmenting the high-resolution anatomical images and down-gridding the obtained white matter masks to the same resolution as the functional data. These nuisance covariates regress out fluctuations unlikely to be relevant to neuronal activity[37]. CC maps were then converted to z-score maps using an AFNI built-in function.

**Supplement iii: Optimization of the Objective Function for Detecting Subnetworks**

The objective function (1) is

$$\underset{K,\{G_k\}}{\operatorname{argmax}} \sum_{k=1}^{K} \exp\{\log(\sum(w_{i,j} \mid e_{i,j} \in G_k)) - \lambda_0 \log(|E_k|)\},$$

where $\lambda_0$ is between 0 and 1, often we choose $\lambda_0 = 1/2$. We solve the above objective in two steps. First, we fix the number $K$. Then

$$\begin{aligned}
&\underset{\{G_k\}}{\operatorname{argmax}} \sum_{k=1}^{K} \exp\{\log(\sum(w_{i,j} \mid e_{i,j} \in G_k)) - \lambda_0 \log(|E_k|)\} \\
&= \underset{\{G_k\}}{\operatorname{argmax}} \sum_{k=1}^{K} \left( \frac{\sum(w_{i,j} \mid e_{i,j} \in G_k)}{|E_k|} \right)^{\lambda_0} \left( \sum(w_{i,j} \mid e_{i,j} \in G_k) \right)^{1-\lambda_0} \\
&= \underset{\{G_k\}}{\operatorname{argmax}} \sum_{k=1}^{K} \rho_{kk} |V_k|, \text{ when } \lambda_0 = 1/2, \rho_{kk} = \sum(w_{i,j} \mid e_{i,j} \in G_k)/|E_k| \\
&= \underset{\{G_k\}}{\operatorname{argmax}} \sum(w_{i,j} \mid e_{i,j} \in G)/|V| - \sum_{k=1}^{k} \sum_{k' \neq k} \rho_{kk'}(|V_k| + |V_{k'}|) \qquad (2) \\
&= \underset{\{G_k\}}{\operatorname{argmax}} \sum_{k=1}^{K} \sum_{k' \neq k} \rho_{kk'}(|V_k| + |V_{k'}|) \\
&= \underset{\{G_k\}}{\operatorname{argmax}} \sum_{k=1}^{K} \sum_{k' \neq k} \frac{\sum(w_{i,j} \mid i \in G_k, j \in G_{k'})}{|V_k||V_{k'}|}(|V_k| + |V_{k'}|) \\
&= \underset{\{G_k\}}{\operatorname{argmax}} \sum_{k=1}^{K} \frac{\sum(w_{i,j} \mid i \in G_k, j \notin G_k)}{|V_k|}
\end{aligned}$$

We solve the objective function (2) by using spectral clustering algorithm RatioCut[13].

Next, we select $K$ by grid searching that maximizes the criteria:

$$\sum_{k=1}^{K} \left( \frac{\sum(w_{i,j} \mid e_{i,j} \in G_k)}{|E_k|} \right)^{\lambda_0} \left( \sum(w_{i,j} \mid e_{i,j} \in G_k) \right)^{1-\lambda_0} \quad (3)$$

The choice of $K$ could have a major impact on the objective function. For instance, when $K = |V|$ all edges are excluded and when $K = 1$ more edges are included within the selected subgraphs. At this step, a larger $\lambda_0$ often leads to detected subnetworks with higher proportion of more informative edges and smaller sizes whereas a smaller $\lambda_0$ often produces larger networks including more informative edges in $G$ . We choose the default choice of $\lambda_0$ as 0.5 because we try to include most differentially expressed edges in subnetworks of 'high quality' (that includes a high proportion of differentially expressed edges). Users could choose $\lambda_0$ based on their needs, for instance, ones who wish to obtain 'high quality' subnetworks could tune $\lambda_0$ towards 1 and optimize K in (3) by using $\lambda_0$ around 0.6. We could also tune $\lambda_0$ by cross-validation. Based on our empirical data analysis, $\lambda_0$ ranges 0.4 to 0.7.

The optimization of the above objective function can be implemented by the algorithms in reference 13. When a more complex subgraph topological structure of $G_k$ (e.g. bipartite subgraph) exists in $G$ instead of the default clique structure, advanced graph topology detection tools are needed (e.g. reference 14). The detected organized subnetworks (with more complex graph topological structures) can increase the objective function (2) as the quality term increases and quantity term is almost unchanged. Therefore, the refined graph topological structure detection algorithms, for instance, k-partite, rich-club, and overlapped subgraphs could further assist to optimize the objective function. In future, more graph topological structure automatic detection tools will be developed, which will be compatible with the objective function (2).

| | | | | | | |
|---|---|---|---|---|---|---|
| 17 | ROL.L | -47.16 | -8.48 | 13.95 | 2 | 'Rolandic Oper L' |
| 29 | INS.L | -35.13 | 6.65 | 3.44 | 2 | 'Insula L' |
| 33 | DCG.L | -5.48 | -14.92 | 41.57 | 2 | 'Cingulum Mid L' |
| 69 | PCL.L | -7.63 | -25.36 | 70.07 | 2 | 'Paracentral Lobule L' |
| 70 | PCL.R | 7.48 | -31.59 | 68.09 | 2 | 'Paracentral Lobule R' |
| 18 | ROL.R | 52.65 | -6.25 | 14.63 | 3 | 'Rolandic Oper R' |
| 19 | SMA.L | -5.32 | 4.85 | 61.38 | 3 | 'Supp Motor Area L' |
| 30 | INS.R | 39.02 | 6.25 | 2.08 | 3 | 'Insula R' |
| 54 | IOG.R | 38.16 | -81.99 | -7.61 | 3 | 'Occipital Inf R' |
| 61 | IPL.L | -42.8 | -45.82 | 46.74 | 3 | 'Parietal Inf L' |
| 63 | SMG.L | -55.79 | -33.64 | 30.45 | 3 | 'SupraMarginal L' |
| 79 | HES.L | -41.99 | -18.88 | 9.98 | 3 | Heschl L' |
| 81 | STG.L | -53.16 | -20.68 | 7.13 | 3 | 'Temporal Sup L' |
| 89 | ITG.L | -49.77 | -28.05 | -23.17 | 3 | 'Temporal Inf L' |
| 90 | ITG.R | 53.69 | -31.07 | -22.32 | 3 | 'Temporal Inf R' |